\documentclass[aps,prd,superscriptaddress,twoside,twocolumn,nofootinbib,%
10pt,showpacs,floatfix]{revtex4-1}

\usepackage{amsmath,amssymb}
\usepackage{graphicx,bm}
\usepackage{slashed}
\usepackage{dcolumn}

\allowdisplaybreaks

\begin{document}


\title{Possible signatures for tetraquarks from the decays of $a_0(980)$, $a_0(1450)$}


\author{K. S. Kim}
\affiliation{School of Liberal Arts and Science, Korea Aerospace University, Goyang, 412-791, Korea}

\author{Hungchong Kim}%
\email{Corresponding author; hungchong@kau.ac.kr}
\affiliation{Research Institute of Basic Science, Korea Aerospace University, Goyang, 412-791, Korea}

\date{\today}


\begin{abstract}

Based on the recent proposal for the tetraquarks with the mixing scheme,
we investigate fall-apart decays of $a_0(980), a_0(1450)$ into two lowest-lying mesons.
This mixing scheme suggests that $a_0(980)$ and $a_0(1450)$
are the tetraquarks with the mixtures of two spin configurations of diquark and antidiquark.
Due to the relative sign differences in the mixtures,
the couplings of fall-apart decays into two mesons are strongly enhanced for $a_0(980)$ but suppressed
for $a_0(1450)$. We report that this expectation is supported by their experimental decays.
In particular, the ratios of the associated partial decay widths, which depend on some kinematical factors and the couplings,
are found to be around
$\Gamma [a_0(980)\rightarrow \pi \eta]/\Gamma [a_0(1450)\rightarrow \pi \eta] = 2.51-2.54$,
$\Gamma [a_0(980)\rightarrow K\bar{K}]/\Gamma [a_0(1450)\rightarrow K\bar{K}] = 0.52-0.89$,
which seems to agree with the experimental ratios reasonably well.
This agreement can be interpreted as the tetraquark signatures for $a_0(980), a_0(1450)$.

\end{abstract}


\maketitle

\section{Introduction}

Tetraquarks have been anticipated for a long time in hadron physics simply because
the quark model~\cite{GellMann:1964nj} does not rule out their existence.
Recent studies on tetraquarks focus mainly on hadrons containing heavy quarks
because possible flavor structures are simpler than the light-quark system.
In the hidden-charmed sector, the tetraquark candidates under active investigations are
$X(3872)$,
$X(3823)$,
$X(3900)$,
$X(3940)$~\cite{Belle03,Aubert:2004zr, Choi:2011fc, Aaij:2013zoa, Bhardwaj:2013rmw, Xiao:2013iha, Abe:2007jna},
and they are expected to have the flavor structure of diquark-antidiquark form,
$cq\bar{c}\bar{q}~ (q=u,d)$~\cite{Maiani:2004vq,Maiani:2014aja,Kim:2016tys}.
In the open-charmed and bottomed sector, the tetraquark possibility
was also investigated in the excited states of $D$ or $B$-mesons~\cite{Kim:2014ywa}
with the flavor structure, $cq{\bar q}{\bar q}, (q=u,d,s)$.

For the light-quark system composed of $u,d,s$ quarks,
possible tetraquark structures are more diverse and certain dynamics is necessary in order
to pin down a specific structure that can be physically realizable.
Indeed, in the original proposal made by Jaffe in the 1970s, tetraquarks are
constructed with diquark-antidiquark form, where the diquark
belongs to spin-0, $\bar{\bm{3}}_c$, $\bar{\bm{3}}_f$
because the color-spin interaction is most attractive with this structure~\cite{Jaffe77a,Jaffe77b,Jaffe04,Jaffe:1999ze}.
The tetraquarks in this picture form a nonet in flavor, $\bar{\bm{3}}_f\otimes {\bm{3}}_f={\bm{8}}_f\oplus {\bm{1}}_f$.
The spin structure is $|J,J_{12},J_{34}\rangle=|000\rangle$, where $J$ is the tetraquark spin,
$J_{12}$ the diquark spin, $J_{34}$ the antidiquark spin. The color structure
is constrained to be $|\bm{1}_c,\bar{\bm{3}}_c,\bm{3}_c\rangle$.
This picture is further developed in Refs.~\cite{MPPR04a,EFG09}
even though this model is still confronted with a two-quark picture involving a $P$-wave excitation~\cite{Torn95}.

Tetraquarks can be bound by the color-spin interaction which acts on all the pairs of quarks.
Assuming all the quarks are in an $S$-wave state,
the interaction applies not only to a quark pair either belonging to the diquark or the antidiquark,
but also to other quark pair made up of one quark in the diquark and the other antiquark in the antidiquark.
In this sense, although the spin-0 diquark is the most compact object, it is not clear
whether the tetraquarks formed from the spin-0 diquarks are most stable.
It may be possible that other diquarks can contribute to the formation of stable tetraquarks.

Along this line, we have recently proposed~\cite{Kim:2016dfq} that the spin-1 diquark
with the flavor and color structure of $\bar{\bm{3}}_f, \bm{6}_c$, which is the second most
compact object among all the possible diquarks~\cite{Jaffe:1999ze}, could
be an important ingredient in the formation of tetraquarks.
Specifically, the spin-0 tetraquarks, in a diquark-antidiquark form, can be constructed also
from the spin and color configurations $|0 1 1\rangle$, $|\bm{1}_c,\bm{6}_c,\bar{\bm{6}}_c\rangle$.
The tetraquarks of this type
are found to mix strongly with the ones above,
$|000\rangle$, $|\bm{1}_c,\bar{\bm{3}}_c,\bm{3}_c\rangle$, through
the color-spin interaction~\cite{Kim:2016dfq}.  The physical states can be realized
by the mixtures of $| 0 0 0\rangle$ and $| 011\rangle$
which diagonalize the hyperfine mass matrix coming from the color-spin interaction.
This mixing causes large gaps in hyperfine masses, which seem to match the
mass differences between $a_0 (980)$, $a_0 (1450)$ as well as $K_0^*(800)$, $K_0^*(1430)$.
In fact, this type of the strong mixing with the spin-1 diquark configuration
was also discussed briefly in Ref.~\cite{Jaffe04, Black:1998wt} whose results on mixing is
consistent with ours. But Ref.~\cite{Jaffe04, Black:1998wt} used this strong mixing
to explain the small masses of the lowest-lying states in the $0^+$ channel
without identifying the other states with higher masses.

In this work, we investigate more concrete signatures for tetraquarks particularly
from the decays of $a_0 (980)$, $a_0 (1450)$. If they are tetraquarks,
their decays are expected to be dominated by a fall-apart mechanism~\cite{Jaffe77b} where
its decay proceeds through a recombination of quark and antiquark into two-meson final states.
This mechanism is unique in the multiquark systems and it can be used to study the decay patterns
of tetraquarks as well as pentaquarks~\cite{Lee:2004bsa}.
This is in contrast to the decay of a quark-antiquark system into two mesons
which necessarily requires a creation of a quark-antiquark pair from the vacuum.

What we want to point out is that, in our tetraquarks,
the two spin-configurations, $|000\rangle$ and $|011\rangle$, through the mixing,
add to each other
in making $a_0 (980)$ but they cancel in making $a_0 (1450)$.
When the resonances simply fall apart into two mesons,
this mixing leads to a strong enhancement of the associated couplings for the former and a suppression for the latter.
Such a behavior of the couplings should be reflected in the partial decay widths which can provide experimental comparison.

In the literature, the structure of $a_0 (980)$ and $a_0 (1450)$, especially with respect to their four-quark nature,
has been investigated
in various ways.
Ref.~\cite{Achasov:2009ee} analyzes the Belle~\cite{Uehara:2009cf} data on $\gamma \gamma \rightarrow \pi^0 \eta$ around
$a_0(980)$ and claims that $a_0(980)$ with the four-quark structure is consistent with the data. A similar conclusion
has been drawn from the radiative decay, $\phi \rightarrow \gamma a_0 (980)$~\cite{Achasov:2003cn} as well as from
strong and electromagnetic decays of $a_0 (980)$~\cite{Giacosa:2006rg}.
There are some models with a hybrid type structure for $a_0 (980)$, $a_0 (1450)$.
In Ref.~\cite{Maiani:2006rq}, $a_0 (1450)$ is viewed as tetraquarks mixed with a glueball
while Ref.\cite{Giacosa:2006tf} considers $a_0 (980)$ as mixtures of tetraquarks and quarkonia.
The other approach~\cite{Boglione:2002vv,Wolkanowski:2015lsa,vanBeveren:1986ea,Dudek:2016cru} suggests
that $a_0(980)$ and $a_0(1450)$ can be dynamically generated from a single ${\bar q}q$ state or from coupled-channel meson-meson
scattering.
Our model is different from all these in that $a_0 (980)$ and $a_0 (1450)$ are viewed as the mixtures of
two possible configurations of tetraquarks only.

This paper is organized as follows.
After a brief introduction of the tetraquark wave functions for $a_0 (980)$, $a_0 (1450)$~\cite{Kim:2016dfq},
we examine possible modes of their fall-apart decays.  We then demonstrate
the suppression or enhancement of the couplings related to the fall-apart decays.
The partial decay widths calculated with these couplings will be compared with the experimental data.

\section{Fall-apart decay modes}
\label{sec:fall-apart}

We start by introducing the tetraquark wave functions
for $a_0 (980)$, $a_0 (1450)$ that we have developed in Ref.~\cite{Kim:2016dfq}.
Our tetraquark wave functions are based on a diquark-antidiquark picture schematically of the form,
$q_1 q_2\bar{q}^3 \bar{q}^4$, where $q_1 q_2$
denotes the diquark, and $\bar{q}^3 \bar{q}^4$ denotes the antidiquark.
By combining the diquark in $\bar{\bm{3}}_f$ with antidiqurk in $\bm{3}_f$,
the tetraquarks form a nonet in flavor.
This means that, $a_0 (980)$ and $a_0 (1450)$, being isovector resonances, belong to an octet, ${\bf 8}_f$.
Their members with positive charge, namely $a_0^+ (980)$ and $a_0^+ (1450)$, share the same flavor structure
\begin{eqnarray}
[{\bf 8}_f]^1_2=\frac{1}{2}(su-us)(\bar{d}\bar{s}-\bar{s}\bar{d})
\label{a0_flavor}\ .
\end{eqnarray}

In our mixing scheme, the spin-0 tetraquarks can have two spin configurations,
$|J,J_{12},J_{34} \rangle=|000\rangle_{\bar{\bm{3}}_c,\bm{3}_c}$, $|011\rangle_{\bm{6}_c,\bar{\bm{6}}_c}$,
with the subscripts denoting the colors of the diaquark and antidiquark.
The tetraquark wave functions for $a_0 (980)$, $a_0 (1450)$ are obtained by diagonalizing the hyperfine masses
which give the mixtures of the forms,
\begin{eqnarray}
&&|a_0 (1450)\rangle = -\alpha |000 \rangle_{\bar{\bm{3}}_c,\bm{3}_c} + \beta |011 \rangle_{\bm{6}_c,\bar{\bm{6}}_c} \label{a_1450}\ ,\\
&&|a_0 (980)\rangle = \beta |000 \rangle_{\bar{\bm{3}}_c,\bm{3}_c} + \alpha |011 \rangle_{\bm{6}_c,\bar{\bm{6}}_c} \label{a_980} \ .
\end{eqnarray}
The mixing parameters are fixed to be~\cite{Kim:2016dfq}
\begin{equation}
\alpha=0.817\ ,\quad  \beta=0.577\ .
\label{mixing parameters}
\end{equation}
We notice that the relative sign between $|000\rangle$ and $|011\rangle$ in Eq.~(\ref{a_1450})
is opposite to that in Eq.~(\ref{a_980}).
So the two states, $|000\rangle$ and $|011\rangle$, cancel in making $|a_0 (1450)\rangle$
while they add up in making $|a_0 (980)\rangle$.

When the resonances, $a_0 (980)$ and $a_0 (1450)$, with the form, $q_1 q_2\bar{q}^3 \bar{q}^4$,
decay into two mesons through a fall-apart mechanism, there are two possible ways that quarks and antiquarks can
be recombined.
One possibility, which we call the (13)-(24) decay, is that $q_1 \bar{q}^3$
get together into a meson, and $q_2 \bar{q}^4$ are combined into another meson.
Another possibility, which we call the (14)-(23) decay,
is that $q_1 \bar{q}^4$ and $q_2 \bar{q}^3$ are combined separately to
form the two-meson final state.
But one can readily show that the (13)-(24) decay yields the same decay patterns as the (14)-(23) decay, coinciding
with the intuitive expectation.
Thus, we will consider only the (13)-(24) decay for illustration purposes.

Our present work focuses on the decays into two {\it spin-0} mesons (pseudoscalar)
simply because these channels, in most cases,
are kinematically allowed and they are experimentally accessible for comparison.
Specifically, in the (13)-(24) decay, each pair must be in the state with spin-0 and color singlet.
The flavor part of Eq.~(\ref{a0_flavor}) in this recombination can be written as
\begin{eqnarray}
[{\bf 8}_f]^1_{2} \doteq (s\bar{d})(u\bar{s})-(s\bar{s})(u\bar{d})\ .
\label{flavor}
\end{eqnarray}
If this four-quark state as grouped above simply falls apart into two pseudoscalar mesons,
the right-hand side, guessing from its quark content, corresponds to
the following decay channels with the accompanying numerical factors as
\begin{eqnarray}
\bar{K}^0 K^+ - \frac{1}{\sqrt{3}}\eta^\prime \pi^+ + \sqrt{\frac{2}{3}}\eta \pi^+
\label{decay_channel}\ .
\end{eqnarray}
As one can see from Particle Data Group (PDG)~\cite{PDG16}, the $\bar{K} K$ and $\eta \pi$ channels
exist in the decays of $a_0 (980)$ and $a_0(1450)$. The other channel $\eta^\prime \pi$ also exists but only
in the $a_0(1450)$ decays and its absent in the $a_0 (980)$ decays can be understood from the kinematical reason.
Thus, our tetraquark model seems promising in a sense that the possible fall-apart modes also appear in PDG.

The fall-apart modes, Eq.~(\ref{decay_channel}), provide the open decay channels in our tetraquark wave functions,
Eqs.~(\ref{a_1450}),(\ref{a_980}).  These constitute one important component when the wave
functions are rearranged in the (13)-(24) basis. Namely,
Eq.~(\ref{decay_channel}) can be constructed from the color singlet part from $q_1 \bar{q}^3$ and $q_2 \bar{q}^4$, respectively
[see Eqs.(\ref{color1}),(\ref{color2}) below].  The other component is
the configuration where $q_1 \bar{q}^3$ and $q_2 \bar{q}^4$ are separately combined into a color octet.
Thus, our tetraquarks contain the component of the open decay channels in their wave function
which affect the masses through the expectation value of the color-spin interaction~\cite{Kim:2016dfq},
and the decay couplings through the mixing.
This role driven by the open decay channels is in some sense
consistent with the findings from the coupled-channel analysis where
the resonances are dynamically generated from the meson-meson decay channels
coupled to the quark-antiquark
channels or others~\cite{vanBeveren:1986ea,Dudek:2016cru,Prelovsek:2013cra,Lutz:2005ip,Padmanath:2015era}.
At this moment, it is not easy to clarify the connection between our approach and the coupled-channel
approach but it is interesting to see that 
the both share the similar physical consequence, namely picking up the crucial effect from the open decay channels.

The spin part of our tetraquarks also needs to be recombined into the (13)- and (24)-pairs from
$|000\rangle$ and $|011\rangle$, and
we need to pick up the configuration where both pairs are in the spin-0 state separately.
Technical details are straightforward (see for example Ref.~\cite{Close}) so we present the final expressions only.
From $|000\rangle$ and $|011\rangle$, the components with $J_{13}=J_{24}=0$ are
\begin{eqnarray}
| 0 0 0 \rangle \rightarrow \frac{1}{2} | 0 0\rangle_{13} | 0 0\rangle_{24}\ ;\  | 0 1 1 \rangle \rightarrow \frac{\sqrt{3}}{2}
 | 0 0\rangle_{13} | 0 0\rangle_{24}\ .
\label{spin12}
\end{eqnarray}
Here, our notation $|0 0\rangle_{13}$, for example, denotes that the spin of the (13)-pair is zero and its spin projection is zero.

The color structure of $|000\rangle$ is $|\bm{1}_c,\bar{\bm{3}}_c,\bm{3}_c\rangle$.
Again one needs to recombine this state into the (13)- and (24)-pairs and picks up the color-singlet part
from both pairs. With the use of the tensor notation for $|\bm{1}_c,\bar{\bm{3}}_c,\bm{3}_c\rangle$,
our statement here can be casted into
\begin{eqnarray}
\frac{1}{\sqrt{12}}  \varepsilon_{abd}^{} \ \varepsilon^{aef}
\Big [ q_1^b q_2^d \Big ]
\Big [ \bar{q}^3_e \bar{q}^4_f \Big ]\rightarrow  \frac{1}{\sqrt{3}} \bm{1}_{c13} \bm{1}_{c24}\ .
\label{color1}
\end{eqnarray}
Here, $\bm{1}_{c13}$ [$\bm{1}_{c24}$] denotes that the (13)-pair [(24)-pair] is in the color singlet state.
The color structure of another spin configuration $|011\rangle$ is $|\bm{1}_c,\bm{6}_c,\bar{\bm{6}}_c\rangle$.
Applying the similar prescription, we find the color-singlet part as
\begin{eqnarray}
\frac{1}{\sqrt{96}} \Big [q_1^a q_2^b+q_1^b q_2^a \Big ]
\Big [\bar{q}^3_a \bar{q}^4_b+\bar{q}^3_b \bar{q}^4_a\Big ]
\rightarrow  \sqrt{\frac{2}{3}} \bm{1}_{c13} \bm{1}_{c24}\ .
\label{color2}
\end{eqnarray}

Inserting Eqs.(\ref{decay_channel}),(\ref{spin12}),(\ref{color1}),(\ref{color2})
into Eqs.(\ref{a_1450}),(\ref{a_980}), we obtain our main results, {\it i.e.}, the relative strengths of the couplings of $a_0(980)$ and $a_0(1450)$
to the channels, $\bar{K}^0 K^+$, $\eta \pi^+$, $\eta^\prime \pi^+$. These are listed in Table~\ref{couplings} with
the common overall factor being omitted.
One can check that the SU(3) relations among the couplings are satisfied separately
for $a_0(1450)$ and $a_0(980)$.  For example, the $a^+_0(980) \eta \pi^+$ coupling is $\sqrt{2/3}$ times
of the $a^+_0(980) \bar{K}^0 K^+$ coupling.

From Table~\ref{couplings}, one can clearly see the enhancement of
the couplings of $a_0(980)$ to the two-meson states compared to those of $a_0(1450)$,
namely about a factor of 4.
This enhancement originates from the relative sign differences in our four-quark wave functions, Eqs.~(\ref{a_1450}), (\ref{a_980}),
which are in fact the consequence of the mixing scheme of our tetraquarks~\cite{Kim:2016dfq}.
Therefore, if this enhancement is confirmed from
experimental data, this could be a
clear signature supporting that $a_0(1450)$ and $a_0(980)$ are tetraquarks.

\section{Partial decay widths}
\label{sec:partial width}


\begin{table}
\centering
\begin{tabular}{c|c|c}  \hline\hline
                     &  $a_0^+(1450)$  & $a_0^+(980)$  \\
 \hline
$\bar{K}^0 K^+$      & $-\frac{\alpha}{2\sqrt{3}}+\frac{\beta}{\sqrt{2}}=0.1722$ & $ \frac{\beta}{2\sqrt{3}}+\frac{\alpha}{\sqrt{2}}=0.7441 $  \\[1mm]
$\eta \pi^+$         & $-\frac{\alpha}{3\sqrt{2}}+\frac{\beta}{\sqrt{3}}=0.1406$ & $ \frac{\beta}{3\sqrt{2}}+\frac{\alpha}{\sqrt{3}}=0.6076 $  \\[1mm]
$\eta^\prime \pi^+$  & $\frac{\alpha}{6}-\frac{\beta}{\sqrt{6}}=-0.0994 $ & $-\frac{\beta}{6}-\frac{\alpha}{\sqrt{6}}=-0.4296$  \\[1mm]
\hline\hline
\end{tabular}
\caption{The relative strengths of the couplings of $a_0^+(980)$, $a_0^+(1450)$
to the channels $\bar{K}^0 K^+$, $\eta \pi^+$, $\eta^\prime \pi^+$, are presented here with common overall factor being omitted.
The mixing parameters $\alpha, \beta$ given in Eq.~(\ref{mixing parameters})
have been used in getting the numbers shown. }
\label{couplings}
\end{table}

Our results given in Table~\ref{couplings} can be tested experimentally
from the partial decay widths of $a_0(1450)$, $a_0(980)$.  To focus on the enhancement of the couplings
while eliminating the dependence on the overall factor,
the relevant quantities to consider would be the ratios of the partial decay widths
\begin{eqnarray}
\frac{\Gamma [a_0(980)\rightarrow \pi \eta]} {\Gamma [a_0(1450)\rightarrow \pi \eta] }\ ;\
\frac{\Gamma [a_0(980)\rightarrow K\bar{K}]} {\Gamma [a_0(1450)\rightarrow K\bar{K}] }\ .
\end{eqnarray}
The similar ratio for the decay, $\eta^\prime \pi^+$, cannot be tested due to its kinematical constraint.
To calculate the partial widths, we take the effective Lagrangians involving derivatives~\cite{Black:1999yz} for
corresponding decay channels
\begin{eqnarray}
{\cal L}_{a_0(1450)}&=&0.1722~ g~ \partial_\mu \bar{K}^0 \partial^\mu K^+ a_0^+ (1450) \nonumber \\
&+& 0.1406~ g~ \partial_\mu \eta \partial^\mu \pi^+ a_0^+ (1450)\ ,\\
{\cal L}_{a_0(980)}&=&0.7441~ g~ \partial_\mu \bar{K}^0 \partial^\mu K^+ a_0^+ (980) \nonumber \\
&+& 0.6076~ g~ \partial_\mu \eta \partial^\mu \pi^+ a_0^+ (980) \ ,
\end{eqnarray}
where the numeric factors have been adopted from Table~\ref{couplings}.
Here $g$ denotes the common overall factor.
The partial decay width for each channel can be calculated straightforwardly.
For example, the partial width for $a_0(980)\rightarrow \pi \eta$ is given by
\begin{eqnarray}
\Gamma [m_{a_0}]=\frac{0.6076^2 g^2 p}{32\pi {m_{a_0}^2}} (m_{a_0}^2-m_\pi^2-m_\eta^2)^2\ ,
\label{partial width}
\end{eqnarray}
where $p$ is the momentum of the decay products in the center of mass frame.
Note that the additional kinematical factors, like $p$ and $m_{a_0}^2-m_\pi^2-m_\eta^2$,
increase as the mass gap between the initial and final states in the decay increases.
The formulas for the other partial widths can be obtained similarly.

We note that the $a_0(980)$ mass is just below the $K\bar{K}$ threshold, $\sim 990$ MeV.
So the decay, $a_0(980)\rightarrow K\bar{K}$,
is possible only when the mass distribution around its central mass
broaden by the total decay width is taken into account.
The total width can be included in our calculation of the partial decay width by
taking an average with respect to the mass distribution.

A resonance with decay width is normally represented by the mass distribution
called the Breit-Wigner type.  In our calculation, due to a numerical reason,
we take a different distribution with {\it faster} fall-off away from the central mass
so that the integral in the averaging process converges faster numerically.
Namely, we take the mass distribution with the exponential type,
\begin{eqnarray}
f(M)\sim e^{-(M-M_c)^2 / A^2}~~{\rm with}~A=\frac{\Gamma_{exp}}{2\sqrt{\ln2}}\label{distribution}\ .
\end{eqnarray}
Here $M_c$, $\Gamma_{exp}$ are the central mass and the total decay width of the resonance of concern.
This form keeps the main features of the Breit-Wigner distribution namely
that $f(M)$ has the maximum at $M=M_c$ and the two values of $M$ at the half maximum of $f(M)$
are separated by $\Gamma_{exp}$.
For a general decay process like, $M \rightarrow m_1, m_2$, the partial width averaged over the mass distribution
is calculated as
\begin{eqnarray}
\langle \Gamma (M_c, \Gamma_{exp}) \rangle = \frac{\int^\infty_{m_1+m_2} \Gamma (M) f(M) dM}{\int^\infty_{m_1+m_2} f(M) dM}\ ,
\label{width av}
\end{eqnarray}
once the inputs, $M_c$ and $\Gamma_{exp}$, are given.
According to PDG~\cite{PDG16}, $M_c=1474$ MeV, $\Gamma_{exp}=265$ MeV for $a_0 (1450)$.  For $a_0(980)$,
$M_c=980$ MeV, $\Gamma_{exp}=50-100$ MeV.  So the total width of $a_0(980)$ is quite uncertain.

Our averaging method, Eq.~(\ref{width av}), provides a simple way to include the resonance width and this prescription
can be applied equally to $a_0 (980) \rightarrow \bar{K}K$ as well as to the other decay processes
like $a_0 (1450) \rightarrow \bar{K}K$, $a_0 (1450) \rightarrow \pi \eta$, $a_0 (980) \rightarrow \pi \eta$.
However, for a resonance like $a_0 (980)$ which has two
decay channels with one channel lying above the resonance mass,
the mass distribution can be well described by a Flatt{\'e} distribution~\cite{Flatte:1976xu}.  It seems however that
this distribution contains various parameters that one has to deal with (See for example Table 1 in Ref.~\cite{Baru:2004xg}).
Thus, it is not a simple matter to implement the Flatt{\'e} distribution in all the decay processes of our concern
on an equal footing.
Instead, our distribution given in Eq.~(\ref{distribution}), although it is simple, can
simulate the Flatt{\'e} distribution reasonably well for the narrow resonance, $a_0 (980)$.
Specifically, the Flatt{\'e} distribution for $a_0 (980)$, which was
obtained from more sophisticate models as shown in Fig.6 of Ref.~\cite{Giacosa:2007bn},
has the strong fall-off away from the central mass and the separation at the half maximum seems to be around 50-60 MeV.
Through the averaging process of Eq.~(\ref{width av}), the delicate difference between the two distributions is expected to
give marginal modification on the partial decay width.

The partial decay width, such as $a_0\rightarrow \pi \eta$,
is calculated from the corresponding formula, Eq.~(\ref{partial width}). Folding this with $f(M)$ as given in Eq.~(\ref{width av}), we
obtain the partial width averaged over the mass distribution, $\langle \Gamma \rangle [a_0(980)\rightarrow \pi \eta]$.
Applying the similar prescription to the other partial decays,
we obtain the ratios among them as
\begin{eqnarray}
&&\frac{\!\!\!\langle \Gamma\rangle [a_0(980)\rightarrow \pi \eta]} {\langle \Gamma\rangle [a_0(1450)\rightarrow \pi \eta] } \Bigg |_{\rm theory}= 2.51-2.54
\label{ratio1_the}\ , \\
&&\frac{\!\!\!\langle \Gamma\rangle [a_0(980)\rightarrow K\bar{K}]} {\langle \Gamma\rangle [a_0(1450)\rightarrow K\bar{K}] } \Bigg |_{\rm theory}= 0.52-0.89
\label{ratio2_the}\ .
\end{eqnarray}
The error bars are from the uncertainty in the total decay width of $a_0(980)$.
The sensitivity to this uncertainty is higher in Eq.~(\ref{ratio2_the}) because $\langle \Gamma\rangle [a_0(980)\rightarrow K\bar{K}]$
picks up the contribution mainly from high tail of the mass distribution.
Note that our results contain the factors coming from the strong enhancement of the coupling ratios, about a factor of 4, as well
as the kinematical factors which in fact
reduce the ratios.
Another thing that we want to emphasize is that Eq.~(\ref{ratio1_the})
is independent of the $\eta-\eta^\prime$ mixing as the additional parameter from this mixing cancels in the ratio.

\section{Comparison with experiment data}
\label{sec:comparison}

To compare our theoretical prediction with the experimental data, we now examine the experimental partial decay widths.
Currently the experimental data are rather limited so our prediction cannot be tested accurately.
But still they can be used to verify our prediction at least in a qualitative level.

For $a_0(980)$, PDG~\cite{PDG16} provides a {\it rough} measurement of the partial width for $a_0(980)\rightarrow \pi \eta$ as well as
the branching ratio, $\Gamma [a_0(980)\rightarrow K\bar{K}] / \Gamma [a_0(980)\rightarrow \pi \eta]=0.183$.
According to these measurements, the partial widths of our concern are
\begin{eqnarray}
&&\Gamma [a_0(980)\rightarrow \pi \eta] \approx 60~ {\rm MeV}\label{a980 wid1}\ ,\\
&&\Gamma [a_0(980)\rightarrow K\bar{K}] \approx 10.98~ {\rm MeV}\label{a980 wid2}\ .
\end{eqnarray}

For $a_0(1450)$, the experimental situation is more unclear.
PDG provides 5 decay modes~\footnote{In fact, PDG reports 6 decay modes from $a_0(1450)$.
Since we are considering the charged state $a^+_0(1450)$, one of the decay modes $a_0(1450)\rightarrow \gamma\gamma$
can be excluded in our analysis.}
with their branching ratios which can be used to determine the partial widths of concern, $\Gamma [a_0(1450)\rightarrow \pi \eta]$,
$\Gamma [a_0(1450)\rightarrow K\bar{K}]$, by equating the total width with the sum of the five partial widths.
However, this way of determination may be questionable due to
the poorly known branching ratios, $\Gamma[a_0(1450)\rightarrow \omega \pi\pi]/\Gamma [a_0(1450)\rightarrow \pi \eta]\approx 10.7$~\cite{Baker:2003jh},
$0\leq \Gamma[a_0(1450)\rightarrow a_0(980) \pi\pi]/\Gamma [a_0(1450)\rightarrow \pi \eta]\leq 4.3$~\cite{Anisovich:2001jb}.
Currently PDG quotes these measurements but does not use them in the analysis of $a_0(1450)$.

Alternatively, one can directly use the partial widths determined by Bugg~\cite{Bugg:2008ig}
who reanalyzes the parameters on $a_0 (1450)$ by including the dispersive corrections from four sets of experimental data.
Since the results of Ref.~\cite{Bugg:2008ig} was not quoted in current PDG, it seems that the consensus might have not been reached
on these within experimental community.

Under this circumstance, it will be reasonable to use both sets of the partial decay widths in our analysis,
one set from Bugg~\cite{Bugg:2008ig} and the other set based on the crude branching ratios given in PDG.
Both sets of partial widths are separately given by
\begin{eqnarray}
\begin{array}{lcc}
\text{Partial width} & \text{Bugg(MeV)} & \text{PDG(MeV)}  \\
\hline
\\[-2.8mm]
\Gamma [a_0(1450)\rightarrow \pi \eta]  & 23.7&15.38 \text{--} 20.49  \\[1mm]
\Gamma [a_0(1450)\rightarrow K\bar{K}]  & 17.7&13.53 \text{--} 18.03
\end{array}
\label{a1450 wid}\ .
\end{eqnarray}
The error bars in the PDG set come from the uncertainty in the branching ratio
$0\leq \Gamma[a_0(1450)\rightarrow a_0(980) \pi\pi]/\Gamma [a_0(1450)\rightarrow \pi \eta]\leq 4.3$.

Now, the experimental ratios corresponding to the theoretical estimates,
Eqs.~(\ref{ratio1_the}) and (\ref{ratio2_the}),
can be calculated by combining Eqs.~(\ref{a980 wid1}), (\ref{a980 wid2}) with Eq.~(\ref{a1450 wid}).
Depending on the partial widths in Eq.~(\ref{a1450 wid}),
we obtain two sets of the experimental ratios indicated by ``Bugg'' and ``PDG'' as,
\begin{eqnarray}
\begin{array}{cccc}
\text{Ratio}&\text{Theory} & \text{Bugg} & \text{PDG} \\
\hline
\\[-2.8mm]
\frac{\!\!\!\Gamma [a_0(980)\rightarrow \pi \eta]}{\Gamma [a_0(1450)\rightarrow \pi \eta]}&2.51\text{--} 2.54 & 2.53 &2.93 \text{--} 3.9  \\[2mm]
\frac{\!\!\!\Gamma [a_0(980)\rightarrow K\bar{K}]}{\Gamma [a_0(1450)\rightarrow K\bar{K}]}&0.52\text{--} 0.89 & 0.62 &0.61 \text{--} 0.81
\end{array}
\label{ratio_com}\ .
\end{eqnarray}
Here, the theoretical ratios, Eqs.~(\ref{ratio1_the}) and (\ref{ratio2_the}), have been listed again
for a clear comparison.
It is quite interesting to see that the theoretical ratios match almost perfectly with the experimental
ratios based on Bugg data, suggesting that our tetraquark model with the mixing scheme works very well.
A somewhat puzzling situation occurs in the experimental ratios based on PDG data in comparison with the theoretical ratios.
The value for $\Gamma [a_0(980)\rightarrow K\bar{K}]/\Gamma [a_0(1450)\rightarrow K\bar{K}]$ agrees very well
with its theoretical estimate but
the ratio for $\Gamma [a_0(980)\rightarrow \pi \eta]/\Gamma [a_0(1450)\rightarrow \pi \eta]$ overshoots the theoretical estimate of
Eq.~(\ref{ratio1_the}) by 0.4 or 1.4.  Nevertheless, this still points toward the
enhancement or suppression of the couplings that we have been advocating although the agreement with the PDG ratio is not precise.
Since the enhancement as well as the suppression of the couplings are a unique feature from our tetraquarks with the mixing scheme,
we believe that our findings could be a strong signature for the existence of tetraquarks.

As a further test of our tetraquark model, one can also consider the branching ratios of $a_0(1450)$ and $a_0(980)$
and compare their theoretical estimates with the experimental ones
even though they are not directly related to the enhancement and suppression of the couplings.
In particular, our calculation
of the branching ratios leads to
\begin{eqnarray}
&&\frac{\langle \Gamma\rangle [a_0(1450)\rightarrow K\bar{K}]} {\!\!\!\langle \Gamma\rangle [a_0(1450)\rightarrow \pi \eta] } \Bigg |_{\rm theory}= 1.09
\label{branchratio1_the}\ , \\
&&\frac{\langle \Gamma\rangle [a_0(980)\rightarrow K\bar{K}]} {\!\!\!\langle \Gamma\rangle [a_0(980)\rightarrow \pi \eta] } \Bigg |_{\rm theory}= 0.23-0.38
\label{branchratio2_the}\ .
\end{eqnarray}
Again, the error bar in the second ratio comes from the uncertainty in the total width of $a_0(980)$, $\Gamma_{exp}=50-100$ MeV.
We notice that, in contrast to Eqs.~(\ref{ratio1_the}), (\ref{ratio2_the}), these results can be
affected by the $\eta-\eta^\prime$ mixing.
These ratios are different from the experimental ratios but the degree of the disagreement may not be enough to
reject our claims above.
For $\langle \Gamma\rangle [a_0(1450)\rightarrow K\bar{K}]/\langle \Gamma\rangle [a_0(1450)\rightarrow \pi \eta]$,
the PDG value is 0.88, smaller than its theoretical estimation by 20 \%, and the ratio from Ref.~\cite{Bugg:2008ig} is 0.77,
smaller than its theoretical estimation by 30 \%.
Also for $\langle \Gamma\rangle [a_0(980)\rightarrow K\bar{K}]/\langle \Gamma\rangle [a_0(980)\rightarrow \pi \eta]$,
the theoretical estimation is somewhat larger than the corresponding PDG value of 0.183.

The similar analysis can be applied to $K_0^*(800)$, $K_0^*(1430)$ and their decays.
We also find that the coupling for $K_0^*(800)\rightarrow \pi K$
is enhanced while the coupling for $K_0^*(1430)\rightarrow \pi K$ is suppressed.  This leads to the calculated ratio of the partial widths,
\begin{eqnarray}
\frac{\!\!\!\langle \Gamma\rangle [K_0^*(800)\rightarrow \pi K]} {\langle \Gamma\rangle [K_0^*(1430)\rightarrow \pi K] } \Bigg |_{\rm theory}= 1.76
\label{ratio_the_K}\ ,
\end{eqnarray}
based on our four-quark picture.
Currently the experimental status for $K_0^*$(800) is quite unclear and its decay modes are almost unknown.  So
this result from tetraquarks cannot be tested with the experimental data.

One may argue that our main prediction, namely the enhancement as well as the suppression of the couplings,
can appear in a model without facilitating the tetraquarks.
In particular, it may be possible to obtain a similar consequence from
a different scenario where $a_0 (980)$ is a tetraquark and $a_0 (1450)$ is a two-quark state.
If this situation occurs, the decay mechanism of $a_0 (1450)$ would be very
different from that of $a_0 (980)$. The resonance, $a_0 (1450)$, being a two-quark state, can
not decay through a fall-apart mechanism while $a_0 (980)$ can.
Thus, the decay couplings of $a_0 (1450)$ and $a_0 (980)$ can not be related
as in Table~\ref{couplings}.

\section{Summary}
\label{sec:summary}

In summary, we have investigated possible signatures for the tetraquarks with the mixing scheme.
Based on fall-apart mechanism, we have studied the decays of $a_0(980)$ and $a_0(1450)$ viewed as tetraquarks.
The couplings associated
with the decays of $a_0(980)$ were found to enhance strongly while those related to $a_0(1450)$ are suppressed.
The enhancement and suppression seem to be supported by experimental data of corresponding partial decays.
These results could be possible signatures for tetraquarks.

\acknowledgments

\newblock
The work of H.Kim was supported by Basic Science Research Program through the National Research Foundation of Korea(NRF)
funded by the Ministry of Education(Grant No. 2015R1D1A1A01059529).
The work of K.S.Kim was supported by the National Research Foundation of Korea (Grant No. 2015R1A2A2A01004727).

\end{document}